\documentclass[fleqn,twoside]{article}
\usepackage{espcrc2}

\usepackage{graphicx}
\usepackage{epsfig}
\newcommand{\AmS}{{\protect\the\textfont2
  A\kern-.1667em\lower.5ex\hbox{M}\kern-.125emS}}

\newcommand{\be}{\begin{equation}}
\newcommand{\ee}{\end{equation}}
\newcommand{\beqn}{\begin{eqnarray}}
\newcommand{\eeqn}{\end{eqnarray}}
\newcommand{\eq}[1]{(\ref{#1})}

\title{
\thispagestyle{empty}
\vspace{-17mm}
\rightline
{\small LU-ITP 2002/016, RNCP-Th02016, KANAZAWA-02-19, ITEP-LAT/2002-09~}
\rightline{\small  19 August, 2002~}
\vspace{5mm}
Monopoles, confinement and the photon propagator in QED$_3$ 
}

\author{M.~N.~Chernodub\address{ITEP, B.Cheremushkinskaya 25, Moscow, 117259,
        Russia and \\ 
        Institute for Theoretical Physics, Kanazawa University, 
        Kanazawa 920-1192, Japan}
        \thanks{M.~N.~Ch. is supported by JSPS Fellowship P01023.},
        E.-M.~Ilgenfritz\address[osa]{Research Center for Nuclear Physics,
        Osaka University, Osaka 567-0047, Japan}
        \thanks{E.-M. I. thanks for the support by the Ministry
        of Education, Culture and Science of Japan (Monbu-Kagaku-sho) and
        the hospitality experienced at RCNP.} 
        and A.~Schiller\address{Institut f\"ur Theoretische Physik,
        Universit\"at  Leipzig, D-04109 Leipzig, Germany}
        \thanks{Presented by A. S. at Lattice'02.}
}
       
\begin{document}

\begin{abstract}
\vspace{-2mm}
We study the lattice gauge boson propagator of $3D$ compact QED
in Landau gauge at zero and non-zero temperature.
Non-perturbative effects are reflected by the generation
of a mass $m$, by an anomalous dimension $\alpha$ and by the photon wave
function renormalisation $Z$. These effects can be attributed
to monopoles: they are absent in the propagator of the
regular part of the gauge field.
The r\^ole of Gribov copies is carefully investigated.
\vspace{-6mm}
\end{abstract}

\maketitle

Three--dimensional compact electrodynamics (cQED$_3$) shares
two essential features with QCD, confinement~\cite{Polyakov}
and chiral symmetry breaking~\cite{ChSB}. 
Confinement of electrically charged particles is caused by a 
plasma of monopoles which emerge due to the compactness of the 
gauge field.
Recently we have interpreted the deconfinement phase transition 
in cQED$_{2+1}$ from the monopole point of view and have shown 
that the transition is independent of the strength of external 
fields~\cite{CISPaper1and2}.
In Ref.~\cite{CISLetter} we have demonstrated how the confinement 
property is manifest in the gauge boson propagator of this 
theory and how the propagator changes at the deconfinement 
temperature. We have found that an anomalous dimension $\alpha$ appears 
which modifies the momentum dependence, apart from the generation 
of a mass $m$, which can be well described by Polyakov's 
theory~\cite{Polyakov}. 
We have shown that all nontrivial effects originate from the 
singular fields of the monopoles and disappear with the 
formation of dipoles.

Here we report on an extension of this study to the case of $T = 0$ 
and a careful investigation on the severity of the Gribov copy problem. 
At $T \ne 0$, it is important to realize that more structure functions 
are necessary for a full description. We found that one of the finite 
$T$ propagators is extremely gauge fixing sensitive.
For our lattice study we have adopted the Wilson action,
$S[\theta] = \beta \sum_p \left( 1 - \cos \theta_p \right)$, where
$\theta_p$ is the $U(1)$ field strength tensor corresponding to 
the compact link field $\theta_l$. The lattice coupling $\beta$ is
related to the lattice spacing $a$ and the continuum coupling 
constant $g_3$ of the $3D$ theory, $\beta = 1 \slash (a\, g^2_3)$.
The Landau gauge is defined as the maximum of a functional
${\cal F}=\sum_l \cos\theta^{G}_l$ with respect to gauge transformations 
$G$. For the Monte Carlo algorithm (combining local and global
updates) and the gauge fixing procedure see Ref.~\cite{Chernodub:2002gp}. 
Simulations for $T=0$ have been performed on a $32^3$ lattice, 
those for $T>0$ on a $32^2 \times 8$ lattice.

We have studied the gauge boson propagator in Landau gauge in lattice 
momentum space $\vec k$ where it is defined as $D_{\mu\nu}({\vec p}) 
= \langle \tilde{A}_{ {\vec k},\mu} \tilde{A}_{-{\vec k},\nu} \rangle$
in terms of the Fourier transformed gauge potential 
$\tilde{A}_{{\vec k},\mu}$ 
for $p_{\nu} = (2/a) \sin (2 \pi k_{\nu}/L_{\nu})$. 
The gauge field in lattice position space is
taken as 
$A_{{\vec n}+\frac{1}{2}{\vec \mu},\mu}= \sin \theta_{{\vec n},\mu} /(g_3\,a)$.
The most general tensor structure of $D_{\mu\nu}$ at $T=0$ is given by
\be
  D_{\mu\nu}(\vec p)= P_{\mu\nu}(\vec p) D (p^2)
  + (p_\mu p_\nu)/p^2 \,  F(p^2)/p^2
\ee
with the $3$-dimensional transverse projection operator
$P_{\mu\nu}(\vec p)= \delta_{\mu\nu}- (p_\mu p_\nu)/{p^2}$.
If the Landau gauge is fulfilled exactly, 
$F(p^2) \equiv p_{\mu}~D_{\mu\nu}~p_{\nu} = 0$.
The transverse and longitudinal propagators, $D$ and $F$, are extracted by
projection and are found approximately rotationally invariant or vanishing.

For $T>0$ the propagator lacks $O(3)$ rotational symmetry. Thus,
we have to consider two scalar functions
$D_{T/L}$ using the $2$-dimensional transverse/longitudinal
projection operators $P_{\mu\nu}^{T/L}$: 
\be
  D_{\mu\nu}(\vec p)= P^T_{\mu\nu} D_T
                    + P^L_{\mu\nu} D_L 
           + (p_\mu p_\nu)/{p^2}  \, F/{p^2} .
\ee
These projectors are given as follows  ($i,j=1,2$; 
$\mathbf p^2= p_i p_i$): 
$P^T_{ij}=\delta_{ij}- (p_ip_j)/{\mathbf p^2}$,
$ P^T_{33}=P^T_{3i}=P^T_{i3}=0$ and
$ P^L_{\mu\nu}= P_{\mu\nu}-P^T_{\mu\nu}$.
The functions $D_T(|\mathbf p |,p_3)$ and $D_L(|\mathbf p |,p_3)$ 
can  be extracted from $D_{\mu\nu}$.
In the static limit, $p_3=0$, we have 
$D_L(|\mathbf p |,0) \equiv D_{33}(|\mathbf p |,0)$, 
{\it i.e.} $D_L$ contains only temporal gauge degrees of freedom
directly exhibiting confinement, while $D_T$ contains the spatial 
gauge degrees of freedom.

In order to discuss the form of the propagator from the monopole plasma
point of view, we decompose the gauge fields links into singular 
(monopole) and regular (photon) contributions,
$\theta_{{\vec n},\mu} = \theta_{{\vec n},\mu}^{phot}
+ \theta_{{\vec n},\mu}^{mono} $~\cite{PhMon} (see also
\cite{Chernodub:2002gp}).  
Once the decomposition of the link angles is done, one is free to define the
corresponding propagators in position space, transform to momentum space, 
and to define the decomposition of $D$ or $D_{L/T}$ (and $F$) into monopole, 
photon and mixed contributions.

%ccccccccccccccccccccccccccccccccccccccccccccccccccccccccccccccccccccc
\begin{figure}[!htb]
\vspace{-3mm}
  \begin{center}
%proc \epsfxsize=6.5cm \epsfysize=4.5cm \epsffile{fig1b.eps}  
    \epsfxsize=6.5cm \epsfysize=4.5cm \epsffile{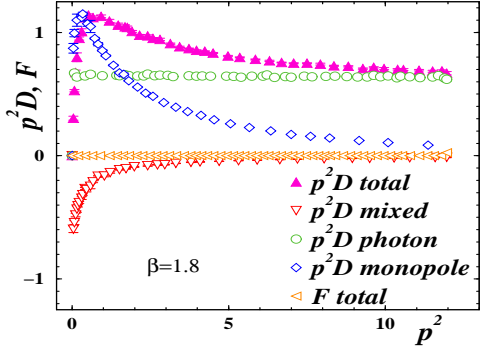}  
    \vspace{-9mm}
    \caption{The propagator $p^2 D$ and its contributions, as well as
             the (vanishing) propagator $F$ vs. $p^2$ at $\beta=1.8$.  
             Only part of the data points averaged over the same $p^2$ 
             are plotted.}   
    \label{fig1}
  \end{center}
  \vspace{-9mm}
\end{figure}
%ccccccccccccccccccccccccccccccccccccccccccccccccccccccccccccccccccccc
In Fig.~\ref{fig1}
we show for $T=0$ at $\beta=1.8$ the measured transverse 
propagator $D$ and its different components in lattice momentum space.
To create the best Landau gauge realization, we have 
evaluated $N_G=20$ Gribov copies, independent maxima of ${\cal F}$ obtained
from maximizing random gauge transforms of the original configuration. 
The monopole part of the transversal propagator $D^{mono}$ reaches a maximum in
the low momentum region before it drops towards $p^2=0$.
The regular (photon) part is singular at $p^2 \to 0$ like $D^{phot} \sim
1/p^2$, while the full transversal propagator is not.
$p^2 D^{phot}$ is flat independently of $\beta$.
As expected, the longitudinal part $F$ vanishes within errors. 

\begin{figure}[!htb]
  \begin{center}
    \vspace{-4mm}
%proc \epsfxsize=6.5cm\epsfysize=4.5cm\epsffile{gc.alpha.DTDL.eps}  
    \epsfxsize=6.5cm\epsfysize=4.5cm\epsffile{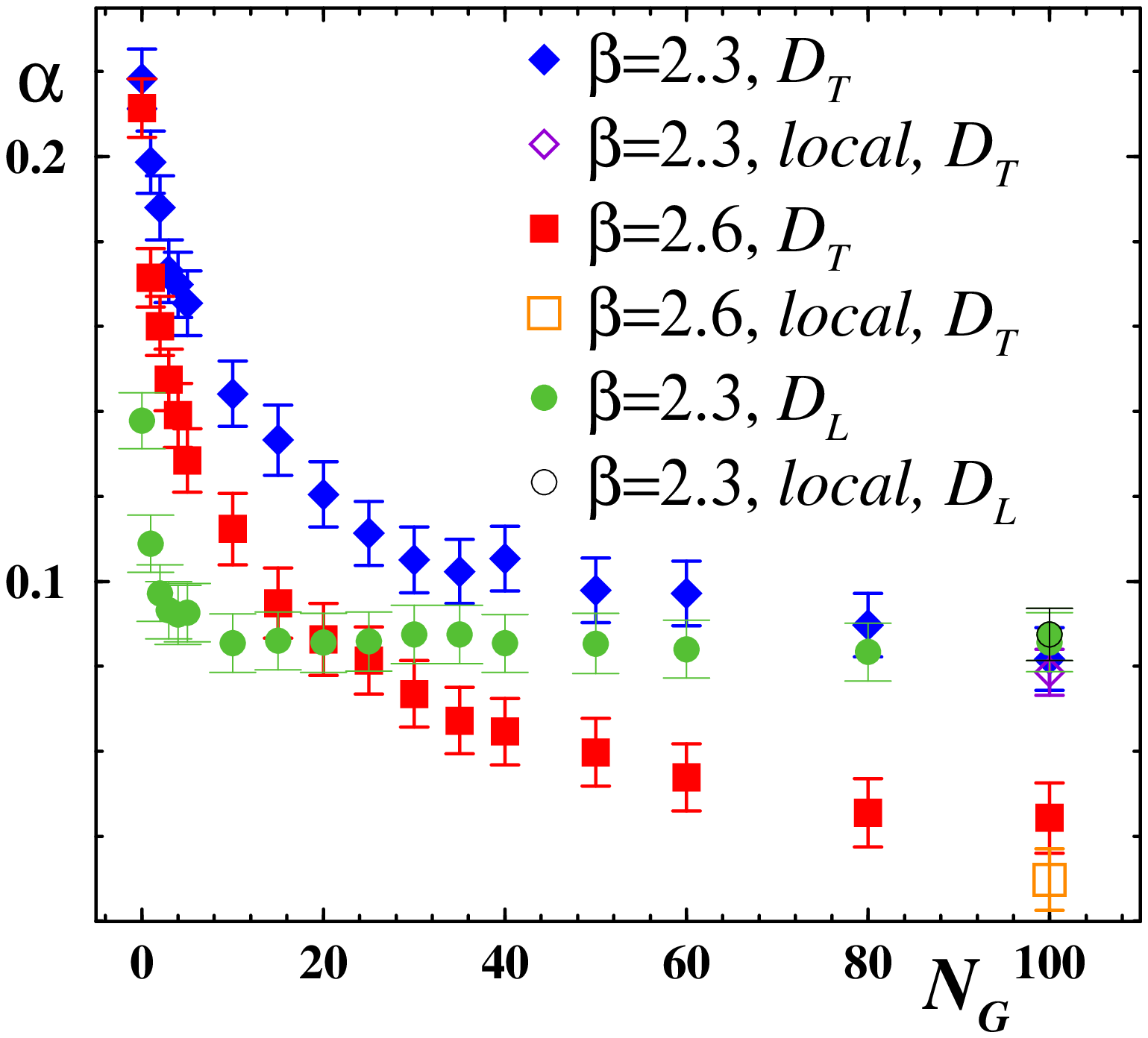}  
    \vspace{2mm}     \\  
%proc \epsfxsize=6.5cm\epsfysize=4.5cm \epsffile{gc.mass.DTDL.eps}    
    \epsfxsize=6.5cm\epsfysize=4.5cm \epsffile{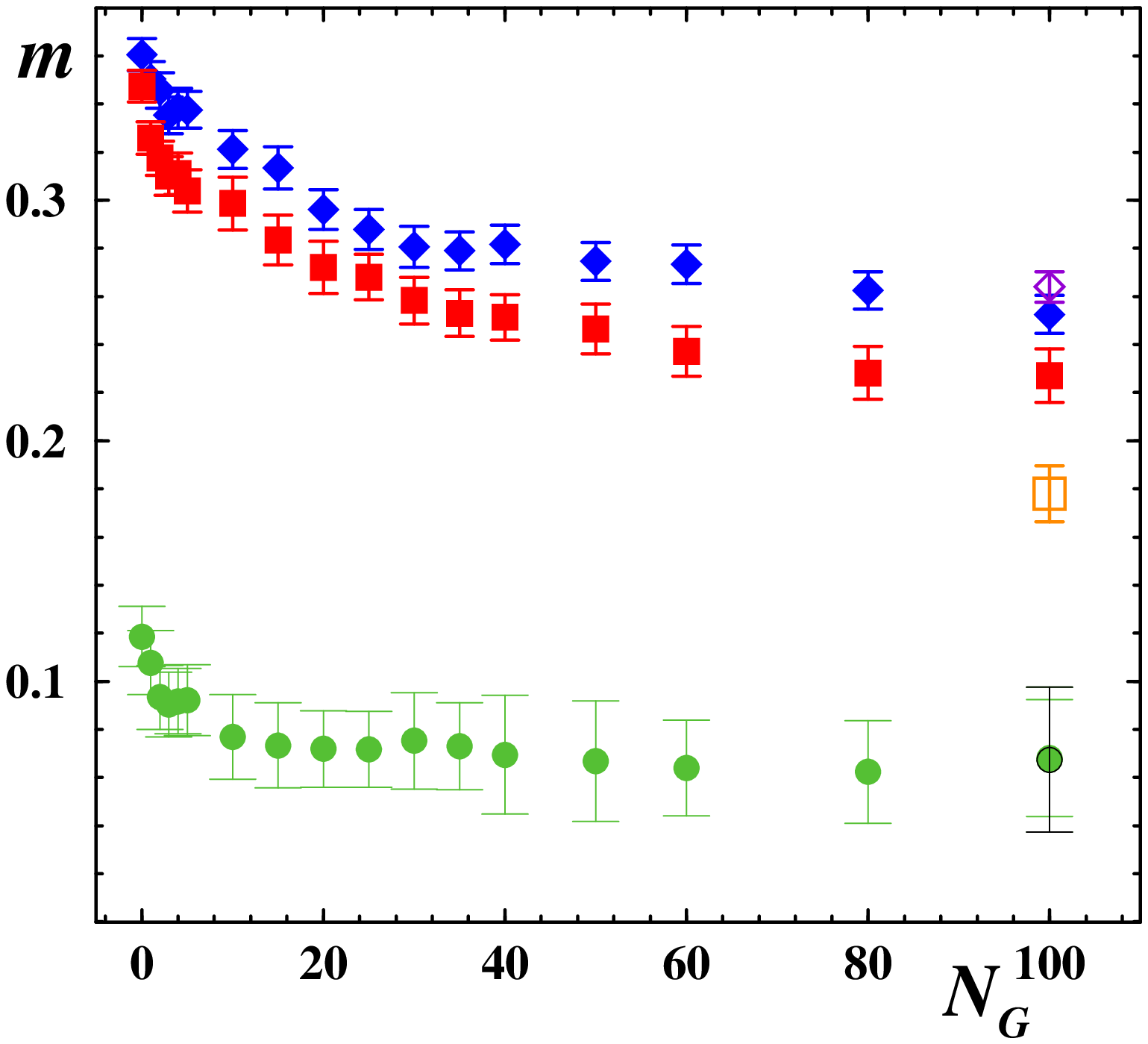}    
    \vspace{-9mm}   
    \caption{Fitted $\alpha$ and $m$ vs. $N_G$ for $D_T$ and $D_L$ at
             different $\beta$'s and updates.    
             For $D_L$ at $\beta=2.6$ we have $\alpha=0$ and $m=0$.}      
    \label{fig2} 
  \end{center}
\end{figure}

Following~\cite{CISLetter} we describe $D$ by the function
\beqn
  D(p^2) = ({Z}/{\beta}) \, {m^{2\alpha}}/[{p^{2(1+\alpha)}+m^{2(1+\alpha)}}]
  + C . \!\!
  \label{def:anomalous_fit}
\eeqn
The photon part $D^{phot}$ is fitted using (\ref{def:anomalous_fit})
with $\alpha = m = 0$.
At finite $T$, the propagator data for $D_L$ and $D_T$
are analyzed for $p_3=0$, as a function of ${\mathbf p}^2$
using the same fit function~\eq{def:anomalous_fit}. 
We have investigated the Gribov copy dependence of $D$ and the finite
$T$ propagators $D_L$ and $D_T$. In this respect, 
the zero temperature $D$ behaves  
similarly as $D_L$. Some results are summarized in Fig.~\ref{fig2}
%ccccccccccccccccccccccccccccccccccccccccccccccccccccccccccccccccccccc
%ccccccccccccccccccccccccccccccccccccccccccccccccccccccccccccccccccccc
at $\beta$ values below and above the phase transition.
After a few Gribov copy attempts $N_G$ all fit parameters of 
$D_L$ are almost insensitive to $N_G$.
The results at large $N_G$ are independent of whether the total Monte 
Carlo algorithm includes global updates or not~\cite{Chernodub:2002gp}  
(see the label ``local '' for the latter case).

The parameters for $D_T$ (and $D_T(|\mathbf p |,0)$ itself), however, 
are strongly dependent on $N_G$ if global updates are included. 
At $N_G=100$ a plateau is not yet reached.
In the deconfinement phase, results are also sensitive to the presence 
of global update steps, in particular at high values of $\beta$.
Runs with only local updates produce propagators with significantly 
lower fit values of $\alpha_T$ and $m_T$, nearer to what is expected. 

\begin{figure}[!htb]
  \begin{center}
%proc \epsfxsize=6.5cm\epsfysize=4.5cm\epsffile{alpha_all.eps}  
     \epsfxsize=6.5cm\epsfysize=4.5cm\epsffile{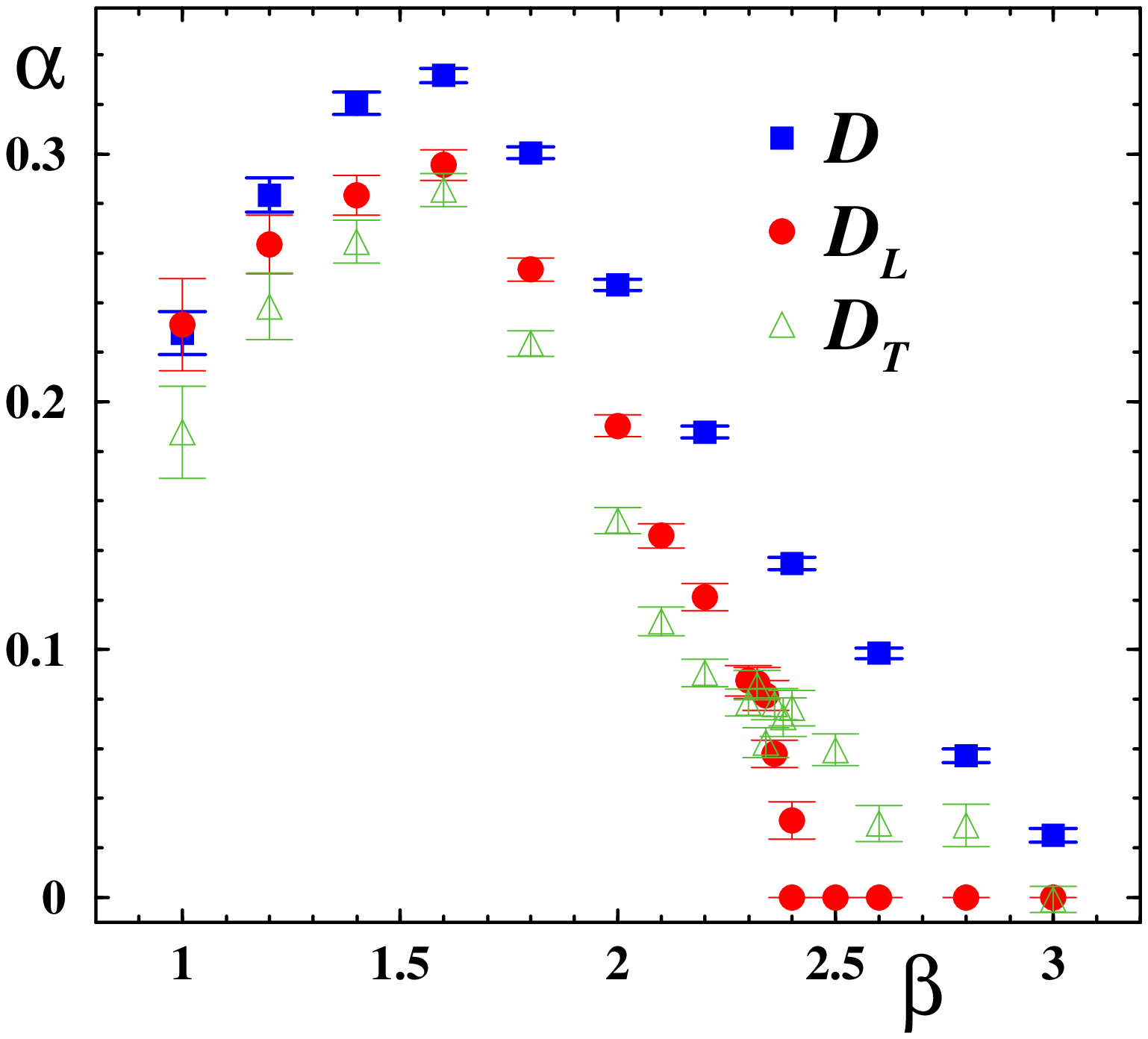}  
     \vspace{3mm} \\        
%proc \epsfxsize=6.5cm\epsfysize=4.5cm\epsffile{mass_all.eps}     
     \epsfxsize=6.5cm\epsfysize=4.5cm\epsffile{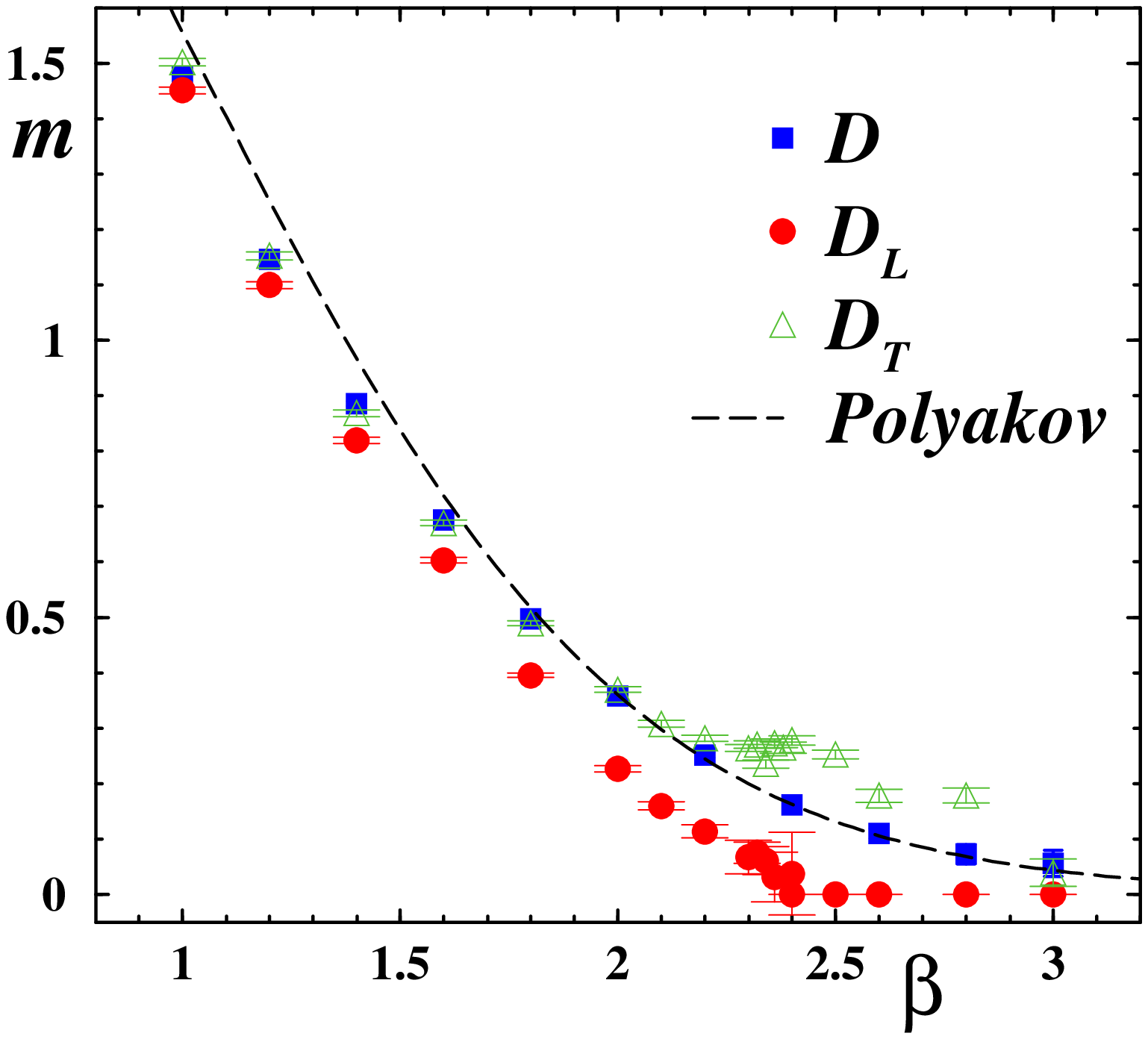}     
     \vspace{-10mm}
     \caption{Fitted $\alpha$ and $m$ vs. $\beta$ for
          $D$, $D_L$, $D_T$.}
     \label{fig3}
  \end{center}
  \vspace{-9mm}
\end{figure}

Realizing this, we decided to use in the final measurements at $T>0$, 
in the region $\beta \ge 2.0$, only local updates and to use not less
than  $N_G=100$ Gribov copies per configuration.
Deep in confinement (below $\beta=2.0$ for $T \ne 0$, and generally for $T=0$)
our standard was $N_G=20$, with global updates included in the Monte Carlo.
Thus, the results for $D_T$ at larger $\beta$ should be considered only 
as qualitative.
The best fit parameters $\alpha$ and $m$ of the 
propagators $D$, $D_L$ and $D_T$ are  presented in Fig.~\ref{fig3}
as functions of $\beta$.
%ccccccccccccccccccccccccccccccccccccccccccccccccccccccccccccccccccccc
%ccccccccccccccccccccccccccccccccccccccccccccccccccccccccccccccccccccc
The anomalous dimensions $\alpha \ne 0$ in the confinement region for all
propagators are functions not only of the monopole density (which is
monotonously decreasing with growing $\beta$). The cluster structure of 
the monopole configurations seems to play a significant role.
The fit parameters $\alpha_L$ and $m_L$ of $D_L$ vanish at $\beta_c$, 
giving a clear signal of the finite temperature phase 
transition~\cite{CISLetter} 
caused by the formation of dipoles. This is not the case for $\alpha_T$ 
and $m_T$ of the $D_T$ propagator. We associate the (so far) inconclusive 
behaviour of $D_T$, at the transition and beyond, with the insufficient 
gauge fixing.
The mass $m$ for $T=0$ is in good agreement with the theoretical 
prediction~\cite{Polyakov} valid for a dilute monopole gas. 

In conclusion, we have studied the gauge boson propagator in cQED$_3$, 
both at zero and non--zero temperatures. We have found that the
propagators in all cases under investigation can be fitted by
(\ref{def:anomalous_fit}) which is a sum of the massive
propagator with an anomalous dimension plus a contact term.
Similar to $D_L$ in the confinement phase~\cite{CISLetter}, 
$D$ for $T = 0$ has an anomalous dimension $\alpha \ne 0$.
The behavior of $D_T$ at the phase transition is obscured by the
extreme sensitivity with respect to Gribov copy effects, {\it i.e.} 
remaining wrapping Dirac strings.

\end{document}